# International collaboration of Ukrainian scholars: Effects of Russia's full-scale invasion of Ukraine

Corresponding author:
Myroslava Hladchenko, Kyiv, Ukraine
hladchenkom@gmail.com

**Abstract**
This study explores the effects of Russia's full-scale invasion of Ukraine on the international collaboration of Ukrainian scholars. First and foremost, Ukrainian scholars deserve respect for continuing to publish despite life-threatening conditions, mental strain, shelling and blackouts. In 2022-2023, universities gained more from international collaboration than the NASU. The percentage of internationally co-authored articles remained unchanged for the NASU, while it increased for universities. In 2023, 40.8 % of articles published by the NASU and 32,2% of articles published by universities were internationally co-authored. However, these figures are still much lower than in developed countries (60-70%). The citation impact of internationally co-authored articles remained statistically unchanged for the NASU but increased for universities. The highest share of internationally co-authored articles published by the NASU in both periods was in the physical sciences and engineering. However, the citation impact of these articles declined in 2022–2023, nearly erasing their previous citation advantage over university publications. Universities consistently outperformed the NASU in the citation impact of internationally co-authored articles in biomedical and health sciences across both periods. International collaboration can help Ukrainian scholars to go through this difficult time. In turn, they can contribute to the strengthening of Europe.

## Introduction

International research collaboration provides access to resources and skills which are required for the creation of new knowledge and technologies (Katz & Hicks, 1997; Katz & Martin, 1997). International collaboration is crucial for any country but even more significant it is for an invaded country like Ukraine.     In 2013 Ukrainians started a Revolution of Dignity because the pro-Russian president rejected to sign an Associate Agreement with the EU. Ukrainians rebelled against exploitation of governmental institutions which implied a high level of corruption and violence of law. Ukrainians have won the Revolution of Dignity. However, the unwillingness of Russia to lose control over Ukraine resulted in the Russian invasion of Ukraine in 2014 (). In 2022, the scope of the Russian invasion extended to all territory of Ukraine.

Continuous shelling and blackouts are an everyday reality for all Ukrainians including scholars. Ukraine regularly faces attacks on its electricity generation system. Not to mention the economic crisis caused by Ukraine allocating all resources to the defence  sector. In 2022, the GDP of Ukraine fell by 29.1%, while the inflation was 26% (BBC, 2024). In 2022, compared to 2021, gross domestic expenditure on research and development in Ukraine decreased by 38.5% from $2019 million to $1241 million (UNESCO, 2024). The war has eroded state budgets allocated to scientific institutions. UNESCO (2024) estimated that US$1.26 billion is needed to restore public research infrastructure in Ukraine damaged by Russians. This covers both the scientific equipment and buildings. All mentioned factors have negative implications for the research capacity of Ukraine (De Rassenfosse et al., 2024).

   According to the State Statistics Service of Ukraine, from 2021 to 2022 the number of scholars in Ukraine decreased by 22% from 68 500 to 53 200. More than one out of ten (12%), namely 10,429 Ukrainian scholars were forced to emigrate or move within the country. Of these, 5,542 (6.3%) were forced to emigrate to other countries, and 4,887 (5.5%) became internally displaced persons. 1 518 Scholars joined the army to defend Ukraine and 70 of them were killed (Cabinet of Ministers of Ukraine, 2024).

Trying to support Ukrainian scholars the EC, national governments and universities provided funded positions and fellowships e.g., MSCA4Ukraine, Volkswagen Foundation, and Aleksander von Humbold Foundation. The US National Academies of Science, in partnership with the Polish Academy of Sciences, launched an initiative to help Ukrainian researchers settle in Poland (OECD, 2022). These initiatives aimed to support scholars who left Ukraine. However, the majority of scholars remain in Ukraine and require support. International collaboration is essential for Ukrainian scholars to go through this tough time.

**Research question and sub-questions**
This study aims to explore the effects of the Russian full-scale invasion of Ukraine on the international collaboration of Ukrainian scholars. The focus lies on scholars affiliated either with the National Academy of Sciences of Ukraine (NASU) or universities. These two sectors employ the major share of Ukrainian scholars. To answer the research question, this paper addresses the following sub-questions comparing internationally co-authored articles published by Ukrainian scholars before (2020-2021) and after (2022-2023) the Russian full-scale invasion of Ukraine:
1. Has the international collaboration of Ukrainian scholars increased in terms of the number of internationally co-authored articles?
2. Have the top ten countries of international collaboration changed?
3. Are there disciplinary differences in the international collaboration of Ukrainian scholars?
4. Has the citation impact of internationally co-authored articles changed?

The data were taken from the CWTS in-house Scopus database and consist of articles published by Ukrainian scholars in 2020-2023.

**System of higher education and science in Ukraine**
After the fall of the Soviet Union in 1991, Ukraine preserved from the Soviet period the division between mainly teaching-oriented higher education institutions and research institutes of the National Academy and Sciences of Ukraine (NASU). This institutional differentiation clearly differed from Western countries, as even in countries with non-university research institutes (such as Germany) universities still performed both tasks equally (Clark, 1983).

National Academy of Sciences of Ukraine was established in 1918. It was created within the establishment of Ukraine as an independent state (Polonska-Vasylenko, 1955). However, in 1919 Ukraine was occupied by Bolsheviks, while in 1922, Ukraine was forcefully annexed by the Soviet Union. Besides the systematic oppression of Ukrainian history, culture and the nation (Polonska-Vasylenko, 1958), preexisting institutions, such as the Ukrainian Academy of Science, which were originally established to foster Ukrainian nation-building, were converted into institutions aimed at promoting Soviet ideology.

After 1991, the preexisting hierarchical governance structure of the academy was preserved, as well as Soviet-style administrators who had long-ties with communist party leadership. They treated science as if no changes had happened in Ukraine. In 1982, 56 % of academicians were the party members. They preserved their positions mainly as directors of research institutes and departments chairs (Josephon & Egorov, 1997). The Board of Academicians (Presidium) runs the academy. The Board consists of 34 academicians. It also has a supporting staff. The funding for the Board is defined in the state budget. In 2020 it was 4.4 million euro (Parliament of Ukraine, 2020).

Academy consists of research institutes that constitute the fourteen disciplinary sections: mathematics; information technology; mechanics; physics and astronomy; earth sciences; material sciences; energy; nuclear physics and energy; chemistry; biology, physiology and molecular biology; general biology; economics; history, philosophy and law; literature, language and art. Physical and technical sciences dominated over biological in the number of researchers and research institutes (Josephon & Egorov, 1997).

The National Academy of Sciences of Ukraine (NASU) receives the largest share of state research funding. In 2023, NASU accounted for 61.2% of the total state research budget (Ministry of Education and Science of Ukraine, 2024). This funding includes both basic institutional support and competitive project-based funding, which is allocated among the structural units of NASU's research institutions.

In the early 1990s, most Ukrainian higher education institutions were renamed into universities. According to the legislation, these relabeled universities were authorised and encouraged to conduct fundamental and applied research. However, 'upgraded' institutions severely lacked common features of universities e.g., research infrastructure and funding (Hladchenko et al, 2020). In addition, academics at universities had a large number of teaching hours, up to 900 hours per year. In 2014, the maximum teaching hours were decreased from 900 to 600 per year. This aimed academics to increase the time for research. According to Ukrainian legislation, academics have a 36-hour working week implying a workload of 1,548 hours per year. This workload must be distributed among teaching, research, methodological work (e.g., writing textbooks), and administrative activities (Parliament of Ukraine, 2002).

Unlike the National Academy of Sciences of Ukraine (NASU), universities do not receive basic state research funding. Instead, they must apply for project-based research grants from the Ministry of Education and Science of Ukraine. The ministry also manages research funding related to Ukraine's participation in the European Union's Horizon 2020 Framework Programme.

Both universities and research institutes of the NASU are eligible apply for project research funding allocated by the National Research Foundation of Ukraine (NRFU). The NRFU was established in 2018 through the reorganisation of the State Fund of Fundamental Research (Cabinet of Ministers of Ukraine, 2018) and began allocating funds to research projects in 2020. It distributes both state funds and funding provided by foreign agencies for joint research projects involving Ukrainian and international researchers. In 2023 Office Horizon Europe Office in Ukraine was established at the NRFU.

After 1991, publication requirements e.g., for doctoral degrees, promotion to professor, primarily focused on articles published in Ukrainian journals. This changed in 2013 when the Ministry of Education and Science introduced a requirement for publications in Ukrainian journals indexed in Scopus and Web of Science (WoS) as part of the doctoral degree criteria. In 2015, publications in Scopus- and WoS-indexed journals became a requirement for the licensing of academic programmes (Cabinet of Ministers of Ukraine, 2015). That same year, they were also added to the requirements for obtaining academic titles such as associate professor and professor (Ministry of Education and Science of Ukraine, 2016; Hladchenko & Westerheijden, 2018). In 2017 publications in journals indexed in these databases became one of the criteria for the research assessment of higher education institutions. The Ministry of Education and Science of Ukraine conducts the research assessment of higher education institutions once in five years (Cabinet Ministers of Ukraine, 2017). In 2019, the government included publications in Scopus- and WoS-indexed journals in the criteria for assessing research projects applying for state funding (Cabinet Ministers of Ukraine, 2019).

To summarise, Ukrainian system of higher education and science largely resides on Soviet institutions and practices. Both national (Scientific Committee of the National Board of Ukraine on Science and Technology, 2023; De Rassenfosse et al., 2023; Bezvershenko &

Kolezhuk, 2022; Gaind & Liverpool, 2023) and international experts (Horizon 2020 Policy Support Facilities, 2017; Schiermeier, 2019) claim that the inability to break away with the Soviet-style institutions, culture and practices holds Ukrainian education and science back and hinders the societal and economic development of the country.

In terms of international collaboration, the findings of OECD (2022) revealed that prior to Russia's invasion of Ukraine in 2014, the major share of internationally co-authored articles was published in co-authorship with Russian scholars. Since 2015, Ukrainian scholars have decreased collaboration with Russian scholars and shifted to collaboration with the scholars from Poland.

**Internationally co-authored articles and hyperauthorship**
International research collaboration is increasing through years (Abramo et al., 2019; Ganzi et al., 2012) as it has manifold benefits: cost-sharing, reduced workloads, access to specialised expertise, and connection with productive academic networks (Wang et al., 2024). International collaboration can be viewed as a consequence of the increasing complexity of science when tacking of problems requires abundant funding, expensive research infrastructure and involvement of multidisciplinary research teams (Adams 2013). Co-authorship is one of the most widely used measures to quantify scientific collaboration in research publications (Batista et al., 2018; Katz & Martin, 1997).

One of the potential benefits of international collaboration is an increase in citations. Prior studies revealed that internationally co-authored papers receive more citations than domestic ones (Persson et al., 2004; Glanzel and Schubert, 2001; Glanzel 2001). First, this can be due to enhanced research quality of internationally co-authored papers (De Moya-Anegon et al., 2018). Second, an increase in the number of authors leads to an increase in impact and this is not just because of self-citations. An increase in affiliation countries and institutions also results in a higher citation impact (Larivière, 2015). Third, the thematic factor influences the impact of internationally co-authored papers (Velez-Estevez et al., 2022). The themes of internationally co-authored papers lie at the forefront of discipline while domestic papers are focused more on well-consolidated themes.

However, apart from international collaboration and the novelty of the topic the other factors also are also correlated with the number of citations to a paper (Tahamtan et al., 2016). This is the quantity and quality of references, overall length of the paper, journal popularity and accessibility (Gargouri et al., 2010; Lawrence, 2001), title (Bowman & Kinnan, 2018), abstract (Martínez & Mammola, 2020), number of authors (Fox et al., 2016) and their reputations, age, gender, race, and academic rank, and whether the research was funded by a grant (Tahamtan et al., 2016). The recent and impactful references published in top-tier journals ensure citations to articles (Mammola et al., 2021; Tahamtan & Bornmann, 2018). Journal impact is also viewed as one of the clearest predictors of future citations (Abramo et al., 2019; Callaham et al., 2002; Stegehuis et al., 2015).

Disciplines differ in term of international collaboration, in particular this refers to the number of authors (Larivière,2015; Cronin et al., 2003; Franceschet & Costantini, 2010). Collaboration is more common in natural sciences than in social sciences (Moody, 2004). Glänzel (2002) found that collaboration is more spread in chemistry and biomedical research than in mathematics. Physics and medicine are two disciplinary areas characterised by hyperauthorship (Cronin, 2001). Hyperauthorship is the phenomenon that takes place in scholarly communication (Cronin, 2001). Hyperauthorship questions the trustworthiness of the scientific communication system and questions the value of authorship itself (Birnholtz, 2006). Rennie et al. (1997) suggested to replace the list of authors with the list of contributors. Though in experimental high-energy physics, the massive scale and cost of conducting path-breaking research necessitate resource concentration around several major laboratories (e.g., CERN,

Fermilab, SLAC), the hyperauthorship masks individual intellectual contribution and raises questions about the nature of authorship and the impact that hyperauthorship has on metrics of scientific achievement. Hyperathorship distorts the research performance measurement at the national and organisational levels (Nogrady, 2023). It can have negative implications for state funding allocation (Mryglod & Mryglod, 2020).

**Materials and methods**
Since 2013 research assessment policies in Ukraine have prioritised publications in Scopus. Accordingly, data for this study were retrieved from the in-house version of the Scopus database at the Centre for Science and Technology Studies (CWTS) at Leiden University.

First, articles published in 2020-2023 with at least one author affiliated with Ukraine were retrieved from a database, resulting in an initial dataset of 61 424 articles. Second, after cleaning the data, articles with foreign affiliations were further divided into internationally co-authored articles (18,072) and articles authored by Ukrainian scholars co-affiliated with foreign institutions (308). The latter group was excluded from further analysis. Fourth, articles were classified based on authorship patterns into three categories: internationally co-authored, domestically co-authored and single-authored. A total of 17,762 articles were authored by scholars affiliated with the National Academy of Sciences of Ukraine (NASU), while 45,586 articles were authored by scholars affiliated with universities. Articles co-authored by scholars affiliated with both the NASU and universities were counted once in each category. Articles co-authored by scholars affiliated with the NASU and universities were counted twice: once in each category.

The disciplinary area of each article was determined using the in-house version of the Scopus database at the Centre for Science and Technology Studies (CWTS). Articles classified by Scopus as multidisciplinary were fractionalised across disciplines.

Field-normalised citation impact (FNCI) was calculated for each article by normalising citation counts by year of publication, discipline and publication type. The percentage of the top 10% most cited globally was used to identify the most impactful publications.

Given the overdispersion of FNCI data, a negative binomial regression was employed to explore the relationships between collaborative status, affiliations, disciplines, and citation impact.

**Results**
**The overview of the dynamics of internationally co-authored articles in 2020-2023**
Figure 1 illustrates the dynamics of the percentage of articles published by the NASU and universities classified by authorship patterns: international collaboration, domestic collaboration and single author. The percentage of internationally co-authored articles is higher in articles published by the NASU than universities. In 2020-2023, it fluctuated around 40% of the total number of articles published by the NASU. The percentage of internationally co-authored articles published by universities increased from 26.2% to 32.3%. This growth implied a fall in the percentage of single-authored and domestically co-authored articles published by universities.

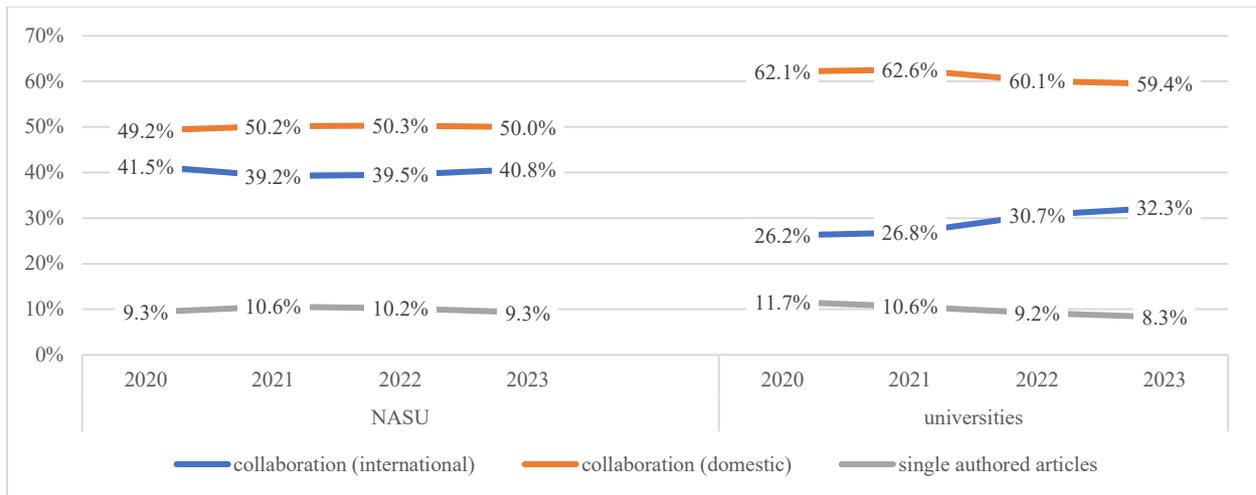

**Fig. 1** Articles 2020-2023 classified by authorship patterns

**Top ten countries of international collaboration**

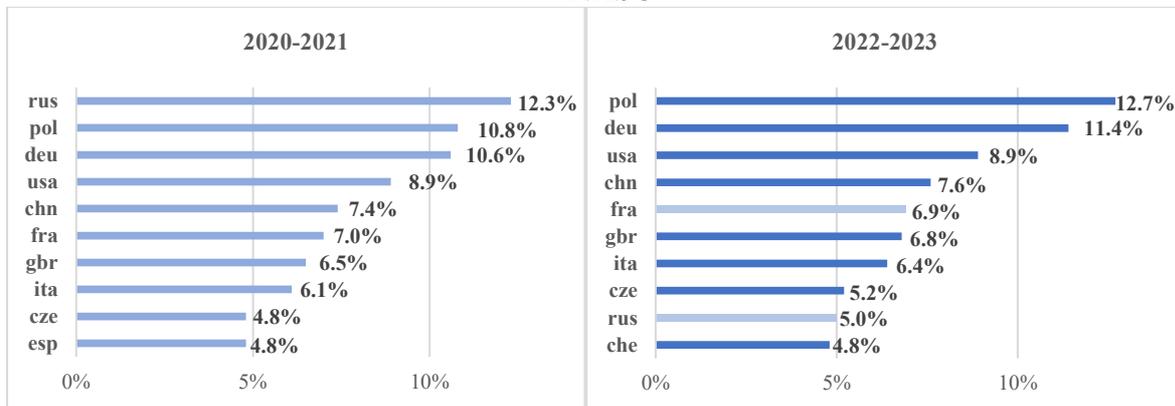

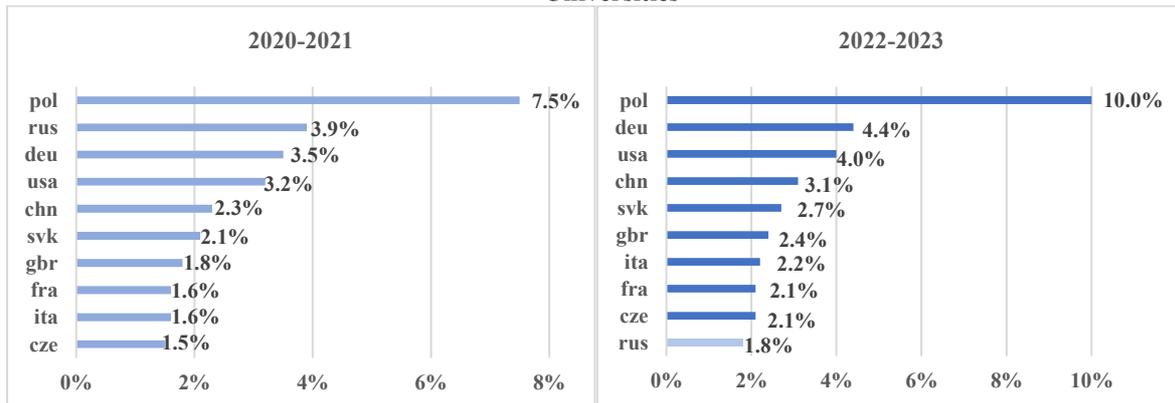

**Fig. 2** Top ten countries of international collaboration (percentage of total articles published by universities).

Figure 2 presents the top ten countries of international collaboration in articles published by the NASU and universities. In 2020-2021, the largest percentage of articles was co-authored with scholars affiliated with Russia, Poland, Germany, USA and China. In 2022-2023, Russia dropped from the first to the ninth position. Most other countries remained consistent across the two periods. while most countries remained the same as in a prior period. The decrease in collaboration with Russia entailed a slight increase in co-authored articles with scholars from Poland, Germany, China, the UK and the Czech Republic.

Top ten countries of international collaboration in articles published by universities largely mirror those in articles published by the NASU, with two notable exceptions. First, Poland leads in both periods. Second, there is a high percentage of articles co-authored with scholars from Slovakia, which rose from the sixth to the second position. All top ten countries show small increases in the percentage of articles published in 2022-2023, with the most substantial growth observed for Poland, rising from 7.5% to 10.0%.

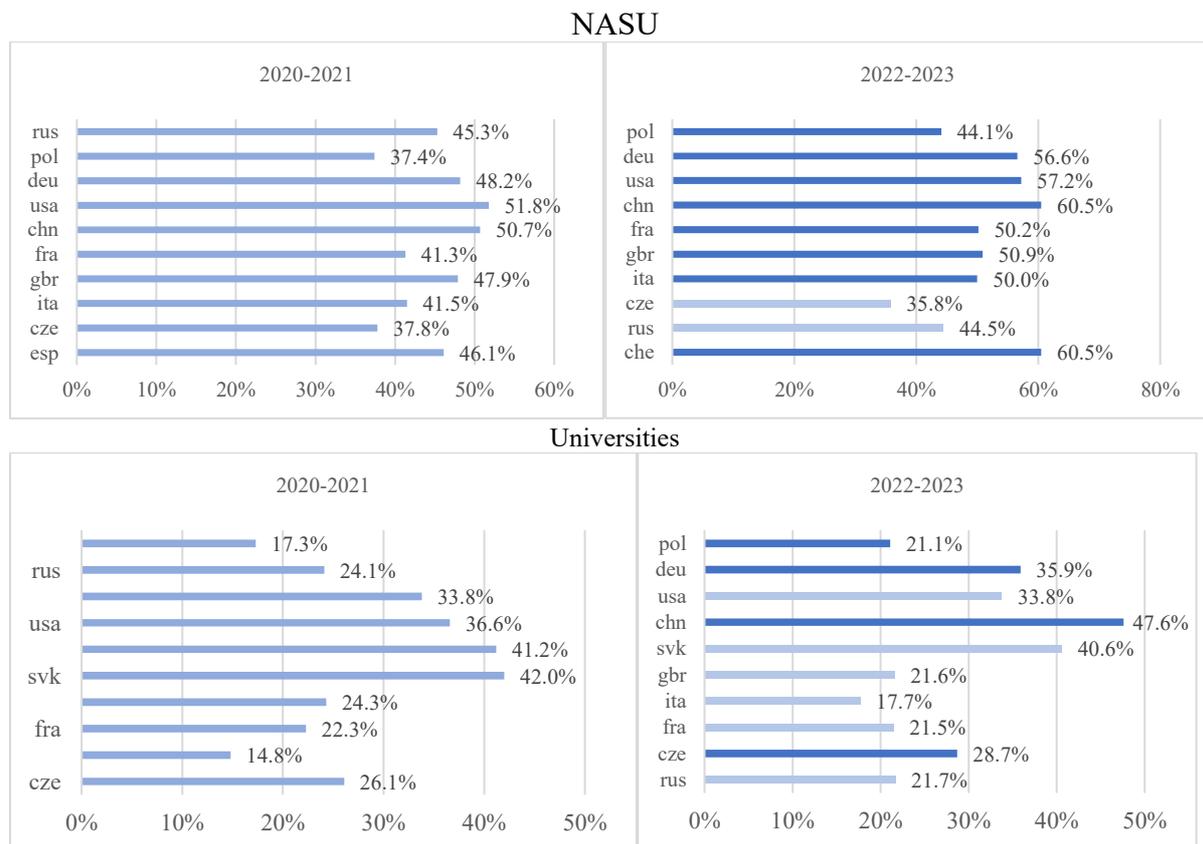

**Fig. 3** The percentage of articles acknowledging the funding of a collaborating country from the top ten countries of international collaboration

Figure 3 illustrates the percentage of articles acknowledging the funding of a collaborating country from the top ten countries in articles published by the NASU and universities. It is higher in articles published by the NASU than in articles published by universities. Arguably this is because the majority of internationally co-authored articles published by the NASU are in physical sciences & engineering which require substantial funding and infrastructure. The highest percentage of articles acknowledging the collaborating country is among articles co-authored with scholars from the USA and China. As regards universities, the highest percentage of articles acknowledging the funding from collaborating country is among articles co-authored with scholars from Slovakia and China. In 2022-2023, the percentage of the articles funded by the ten countries of international collaboration increased for most countries in articles published by NASU and slightly decreased in articles published by universities.

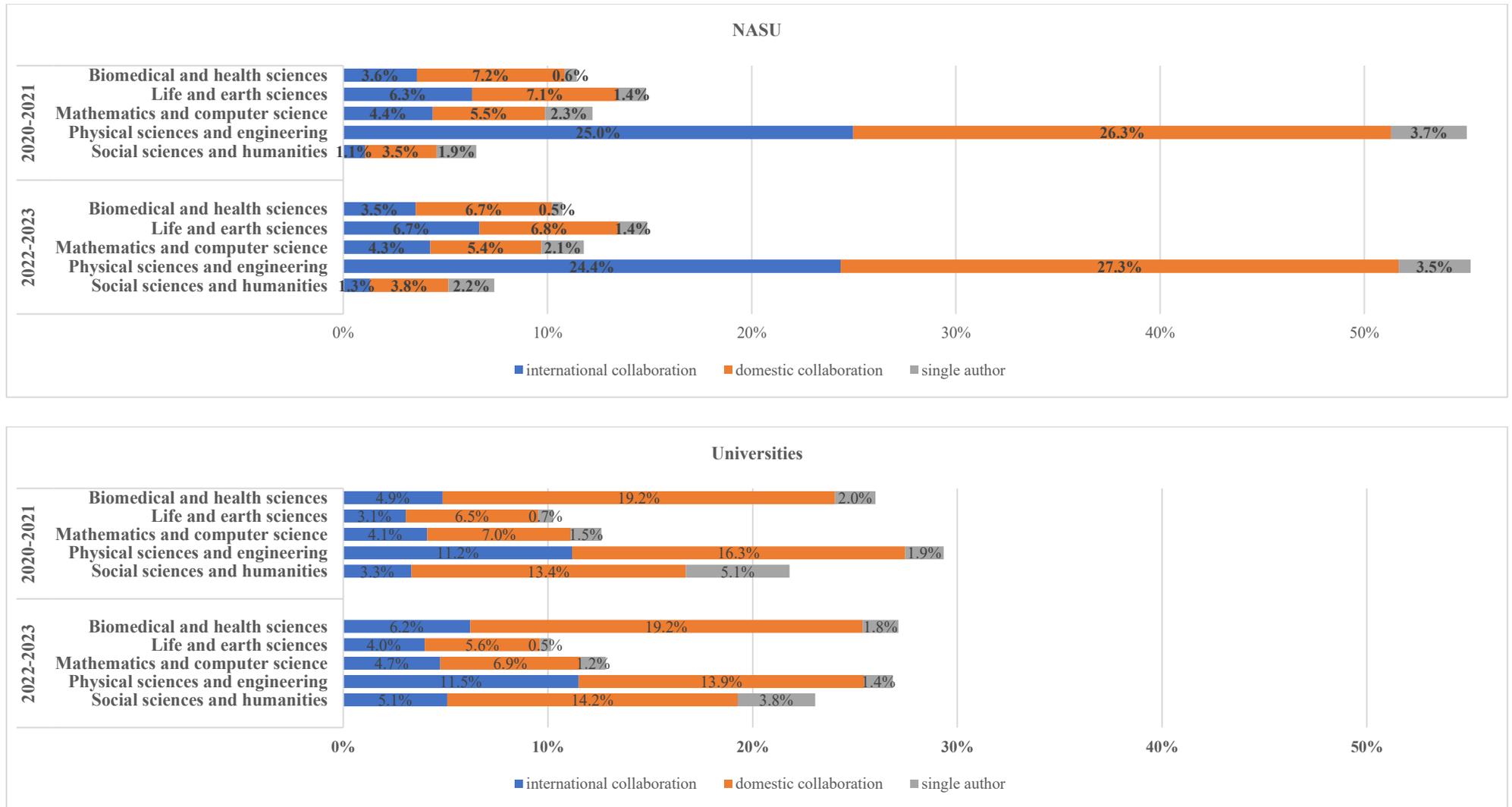

**Fig. 4** Articles 2020-2023 classified by authorship pattern and discipline (universities)

**Authorship patterns across disciplines**
Figure 4 presents an analysis of articles published by the NASU and universities, classified by authorship patterns across disciplines. In both periods, articles in physical sciences & engineering represent the largest share of the NASU's research output, with this share consistently fluctuating around 55.0%. Conversely, social sciences and humanities account for the smallest share of internationally co-authored articles, comprising 1.1% and 1.3% in the respective periods. Notably, domestically co-authored and single-authored articles make up a larger share than internationally co-authored articles across all disciplines. Overall, the distribution of articles across disciplines shows no substantial differences between the two periods.

The research output published by universities is relatively balanced across disciplines. Unlike the NASU, the percentage of articles in physical sciences & engineering does reach 50%. Instead, it accounts for 26.9% of articles published in 2020–2021 and 29.3% in 2022–2023. However, similar to the NASU, the major share of internationally co-authored articles published by universities is in physical sciences & engineering.

**Relationship between the number of foreign affiliation countries and citation impact across disciplines**

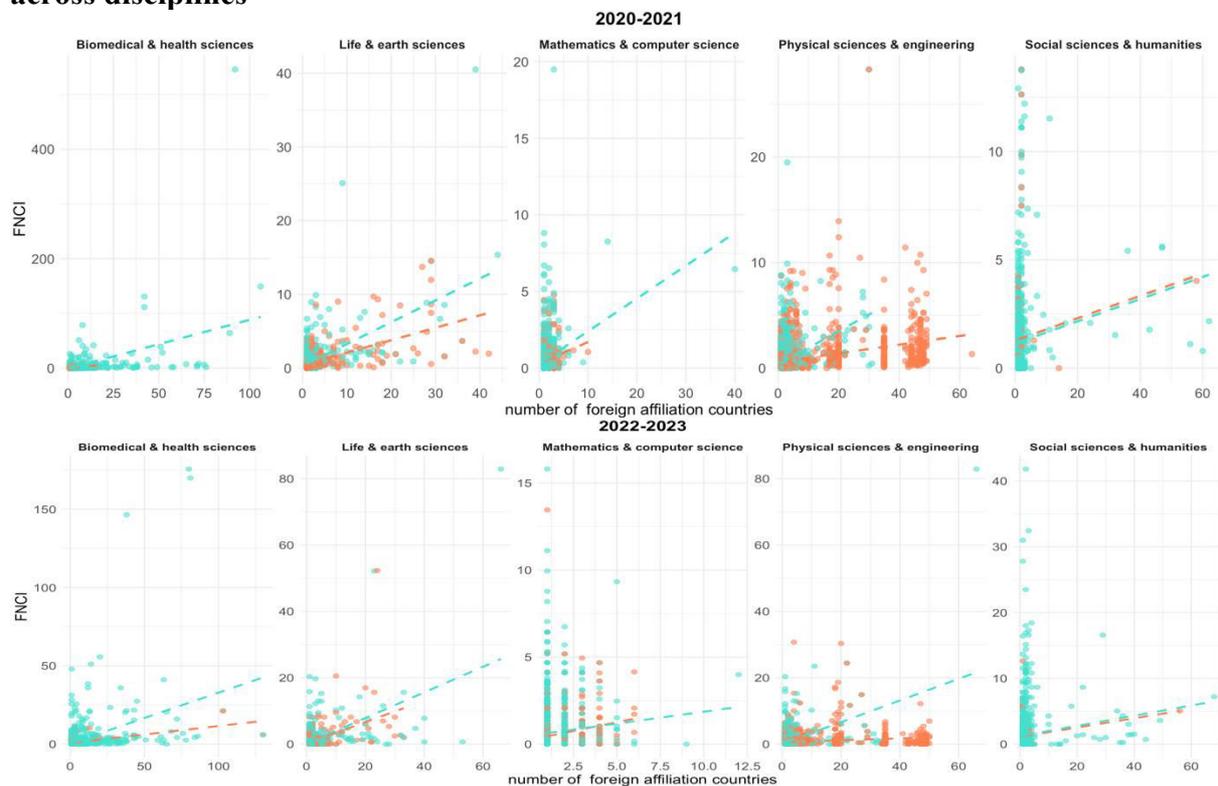

**Fig. 5** Relationship between the number of foreign affiliation countries per article and citation impact in internationally co-authored articles (NASU – orange, universities – greeen)

Figure 5 illustrates the relationship between the number of foreign affiliation countries per article and citation impact (FNCI) across disciplines. It varies significantly across disciplines and between NASU (orange) and universities (green). In most disciplines, the dashed trend line for articles published by universities shows a positive slope, suggesting that an increase in the number of foreign affiliation countries correlates with higher FNCI values. This trend appears stronger in life & earth sciences and biomedical & health sciences. Articles published by the NASU generally display weaker correlations or flatter trend lines compared to those published by universities.

**Relationship between the number of authors and citation impact across disciplines**
Figure 6 illustrates the relationship between the number of authors and the FNCI across disciplines. In both periods, articles published by universities show a stronger positive correlation between the number of authors and FNCI compared to articles published by the NASU. This suggests that universities benefit more from larger collaborations in terms of citation impact.

Biomedical & health sciences and life & earth sciences exhibit a stronger influence of collaboration on FNCI compared to mathematics & computer science or social sciences & humanities. However, this effect is more evident in articles published by universities than by the NASU. In mathematics & computer science most articles irrespective of whether they are published by NASU or universities are concentrated at low FNCI values and authorship counts. In social sciences & humanities, the maximum number of authors has increased from 150 to 500 but collaborations involving a high number of authors have not resulted in noticeable citation impact improvement. Similarly, in physical sciences & engineering, large collaborations involving the NASU do not correspond to higher FNCI values.

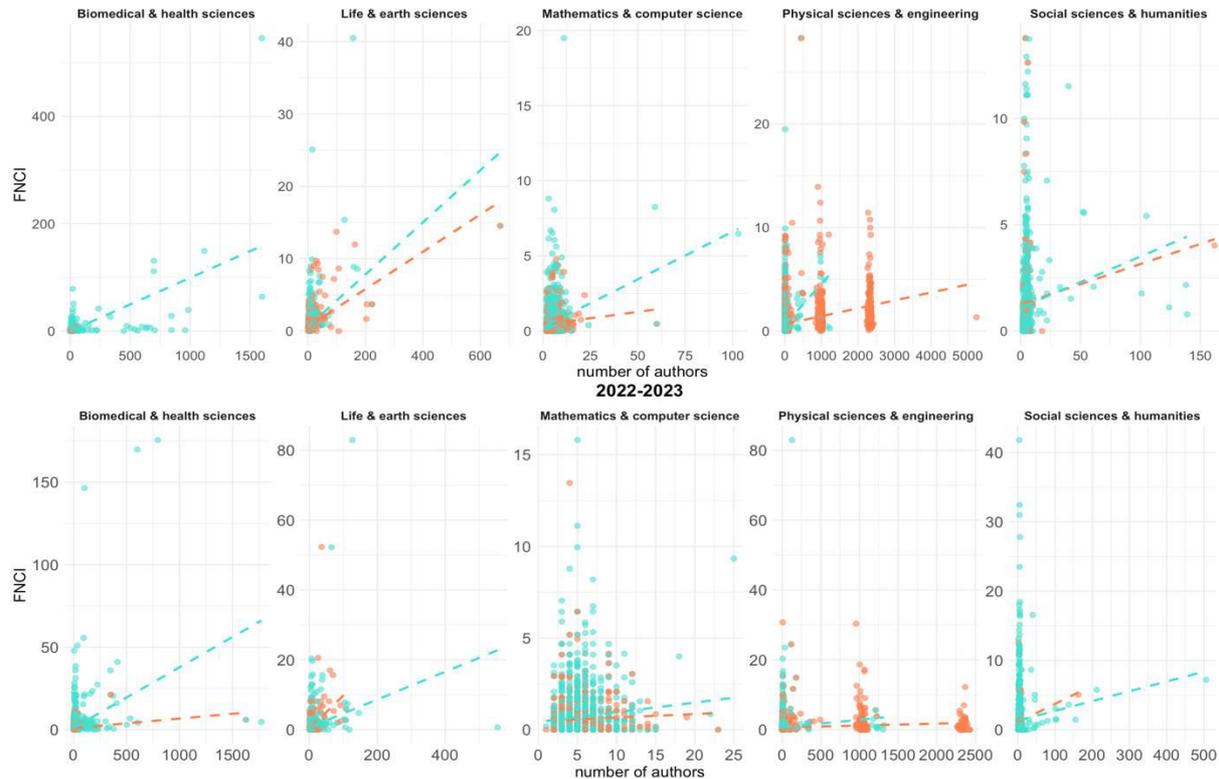

**Fig. 6** Number of authors per article (international collaboration) across disciplines (NASU – orange, universities - green)

**Authorship patterns and citation impact**
Figure 7 illustrates the distribution of FNCI classified by authors' affiliation and authorship pattern. They reveal that internationally co-authored articles have the highest mean FNCI. The mean FNCI of internationally co-authored articles published by the NASU has increased from 0.84 to 0.86, while for universities, it rose from 1.04 to 1.09. The higher mean FNCI of articles published by universities is the result of outliers.

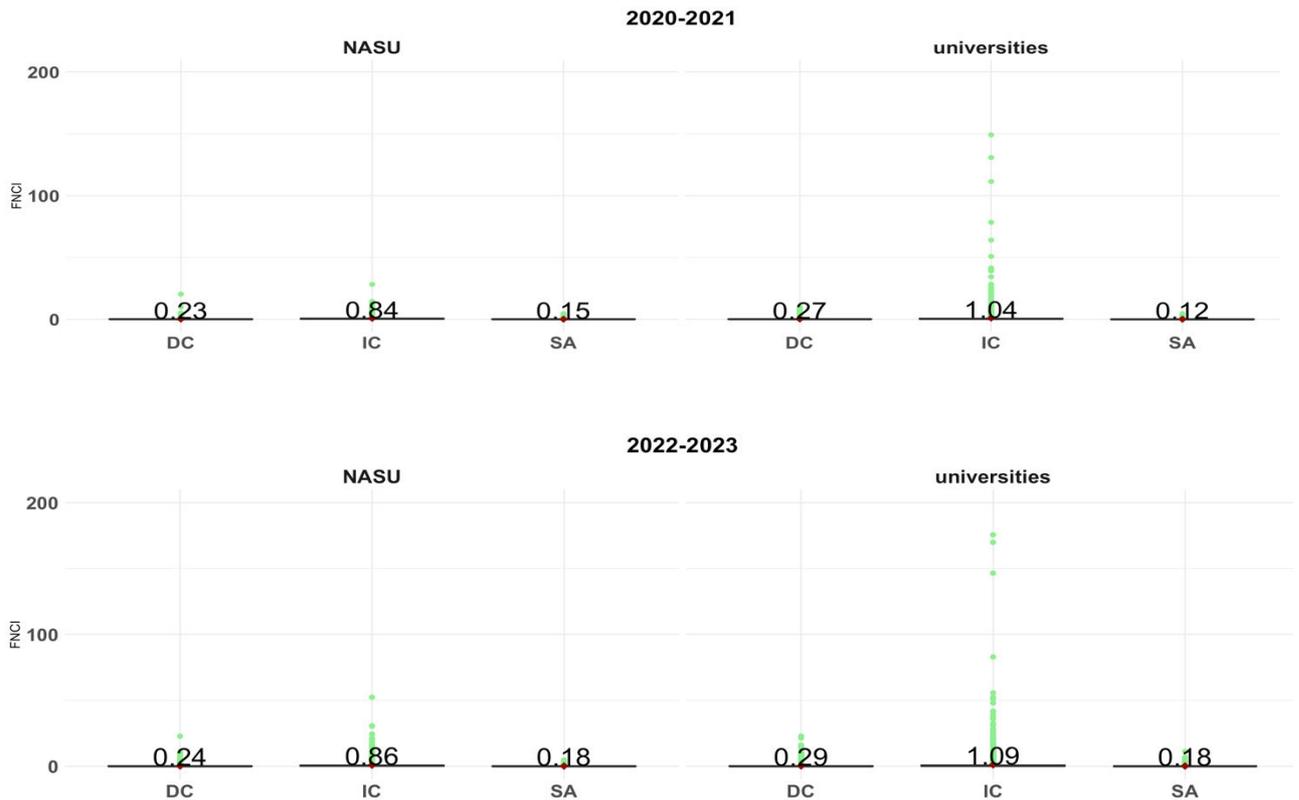

**Fig. 7** Distribution of citations classified by authors' affiliation and authorship pattern (articles published in 2020-2021). DC – domestic collaboration, IC – international collaboration, SA – single author. An article with FNCI equal 549, published in 2020-2021 by a scholar affiliated with the university is not highlighted in Fig 8 to provide clarity in visualisation.

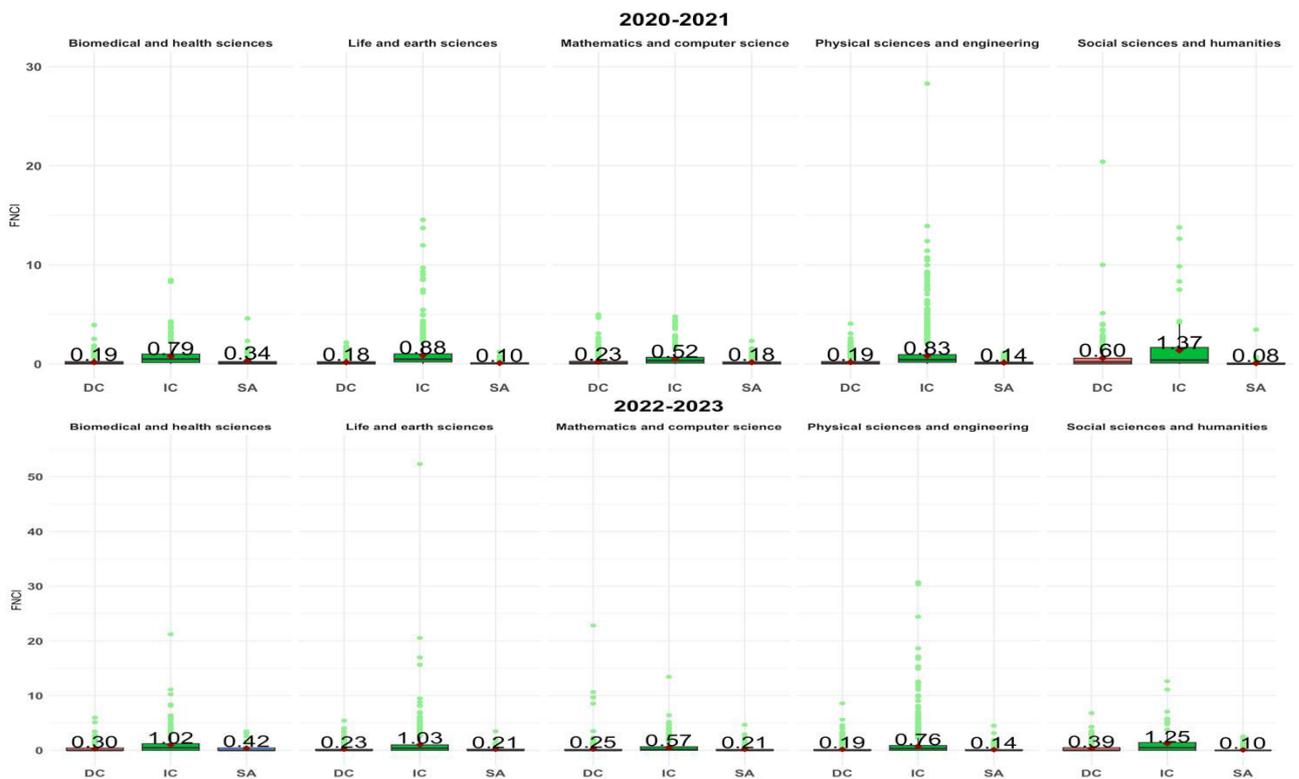

**Fig. 8** Distribution of citations across collaborative status and disciplines (NASU). DC – domestic collaboration, IC – international collaboration, SA – single author.

Figure 8 presents the distribution of citations to articles classified by authorship patterns across disciplines. Figure 13 highlights that internationally co-authored articles in social sciences & humanities (1.37) have the highest mean FNCI among articles published by the NASU in both periods, 1.37 and 1.25 correspondingly. However, the median FNCI in this discipline is much lower, though it has increased to 0.5 in articles published in 2022-2023. The mean FNCI above the world average (1.0) also exhibit articles in biomedical & health sciences and life & earth sciences published in 2022-2023. Conversely, articles in mathematics & computer science have the lowest mean FNCI in both periods. As regards the domestically co-authored articles, those in social sciences & humanities have the highest mean FNCI (0.6 in 2020-2023 and 0.39 in 2022-2023). The highest average FNCI among single-authored articles have articles in biomedical & health sciences (0.34 in 2020-2021 and 0,42 in 2022-2023).

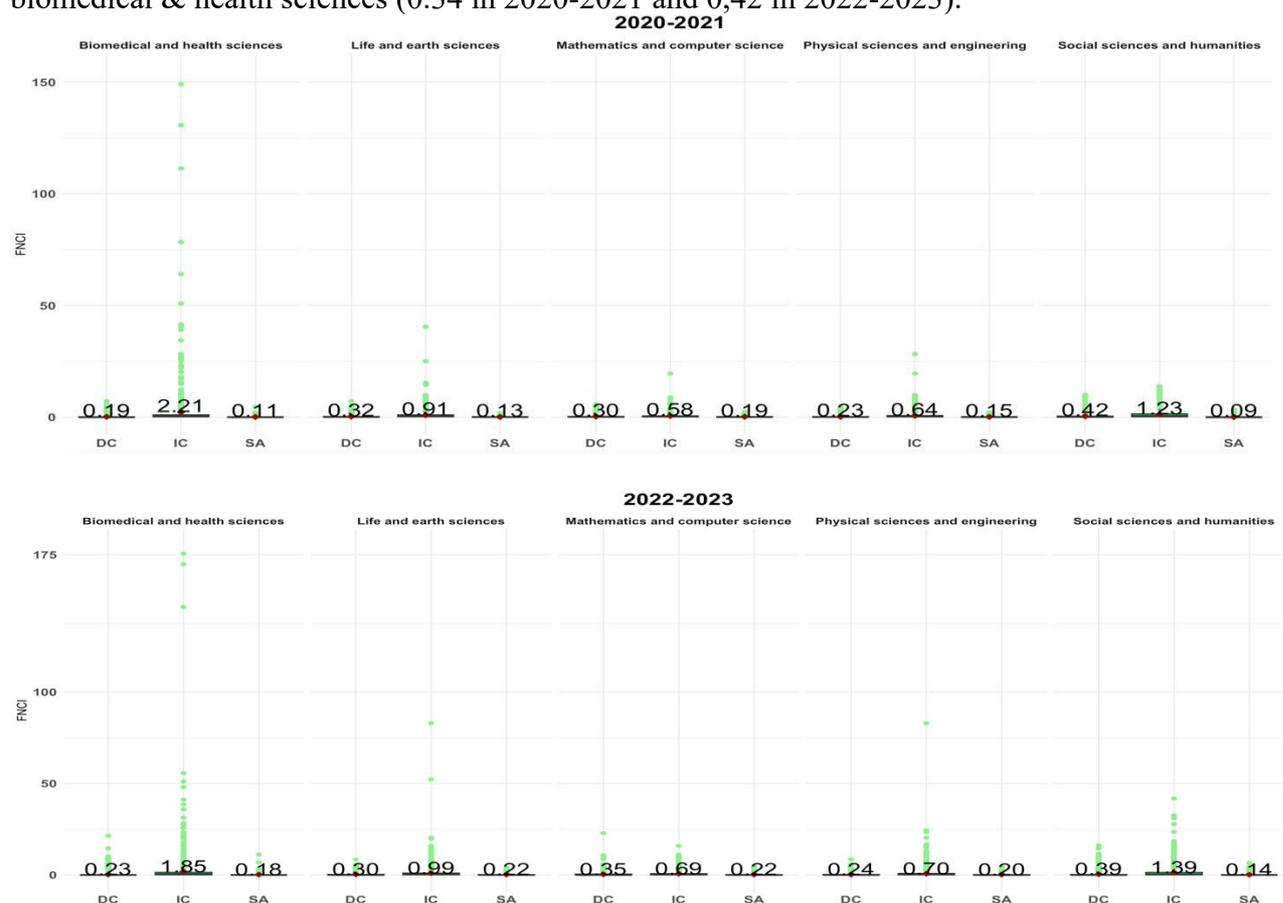

**Fig. 9** Distribution of citations across collaborative status disciplines (universities). DC – domestic collaboration, IC – international collaboration, SA – single author.

Figure 9 displays that in both periods, among the articles published by universities, the mean FNCI above 1.0 have internationally co-authored articles in biomedical & health sciences and social sciences & humanities. This aligns with the findings on the citation impact of articles in these disciplines published by the NASU. However, the high mean FNCI of articles in these disciplines occurs because of outliers, as the median FNCI in all cases does not exceed 0.5. In biomedical & health sciences, the mean FNCI has fallen from 2.21 to 1.35 but the median has risen from 0.38 to 0.45. Conversely, in social sciences & humanities, the mean FNCI has increased from 1.23 to 1.39 but the median has fallen from 0.5 to 0.25. Articles in mathematics & computer science consistently exhibit the lowest mean FNCI among internationally co-authored publications in both periods, followed by articles in physical sciences and engineering. The mean FNCI values of internationally co-authored articles published by the

NASU and univerisities are almost equal across disciplines, except in biomedical & health sciences, where the mean FNCI of articles published by universities is notably higher.

**Statistical tests**

**Table 1** Citation impact of internationally co-authored articles: negative binomial regression

|  | Model 1 NASU (IC) | Model 2 universities (IC) | Model 3 NASU vs universities (IC) 2020-2021 | Model 4 NASU vs universities (IC) 2022-2023 | Model 5 disciplines, NASU, IC,2020-2021 | Model 6 disciplines, NASU, IC,2022-2023 | Model 7 disciplines, universities, IC,2020-2021 | Model 8 disciplines, universities, IC,2022-2023 |
|---|---|---|---|---|---|---|---|---|
| Intercept | -0.17419*** (4.44e-14) | -0.095*** (2.79e-06) | -0.174*** (1.81e-13) | -0.153*** (6.83e-08) | -0.241*** (0.0004) | 0.021 (0.792) | 0.439*** (< 2e-16) | 0.484*** (< 2e-16) |
| Universities |  |  | 0.079** (0.008) | 0.193*** (1.26e-08) |  |  |  |  |
| 2022-2023 | 0.022 (0.513) | 0.136*** (6.23e-07) |  |  |  |  |  |  |
| Life and earth sciences |  |  |  |  | 0.117 (0.159) | 0.004 (0.965) | -0.535*** (< 2e-16) | -0.498*** (2e-16) |
| Mathematics and computer science |  |  |  |  | -0.419*** (6.88e-06) | -0.581*** (3.57e-08) | -0.990*** (< 2e-16) | -0.857*** (2e-16) |
| Physical sciences and engineering |  |  |  |  | 0.058 (0.422) | -0.295*** (0.0004) | -0.884*** (< 2e-16) | -0.845*** 2e-16 () |
| Social sciences and humanities |  |  |  |  | 0.558*** (1.58e-05) | 0.120 (0.180) | -0.228*** (0.0004) | 0.152** (0.010) |
| Dispersion parameter | 1.2925 | 0.6945 | 1.1445 | 0.6243 | 2.2673 | 0.958 | 1.114 | 0.6214 |
| Deviance explained | 5.28208e-05 | 0.0017 | 0.0006 | 0.0029 | 0.0162 | 0.0155 | 0.0608 | 0.0456 |
| McFadden's pseudo-R² | 0.0773 |  |  |  |  |  |  |  |
| Obs. | 7147 | 13201 | 9808 | 10540 | 3687 | 3460 | 6121 | 7080 |

Significance levels:***p< 0.001,**p < 0.01, *p < 0.05

**Table 2** Citation impact across disciplines in internationally co-authored articles: negative binomial regression

|  | Model 1 Biomedical & health sciences | | Model 2 Life & earth sciences | | Model 3 Mathematics & computer science | | Model 4 Physical sciences & engineering | | Model 5 Social sciences & humanities | |
|---|---|---|---|---|---|---|---|---|---|---|
| NASU | -0.241** (0.008) | 0.021 (0.837) | -0.124* (0.0119) | 0.025 (0.679) | 0.660*** (<2e-16) | 0.560*** (<2e-16) | -0.182*** (3.39e-13) | -0.274*** (<2e-16) | 0.318* (0.018) | 0.220 (0.170) |
| universities | 0.680*** (1.17e-11) | 0.464*** (1.97e-05) | 0.029 (0.653) | -0.039 (0.604) | 0.109 (0.138) | 0.187* (0.0178) | -0.263*** (1.80e-13) | -0.086* (0.040) | -0.107 (0.456) | 0.111 (0.507) |
| Dispersion parameter | 0.5817 | 0.4308 | 1.770 | 0.692 | 3.7816 | 1.2958 | 2.4677 | 0.9883 | 0.9685 | 0.4678 |
| Deviance explained | 0.0228 | 0.009 | 0.0001 | 0.0001 | 0.0011 | 0.003 | 0.009 | 0.0007 | 0.0005 | 0.0003 |
| Obs. | 1610 | 1865 | 1707 | 1935 | 1798 | 1886 | 5581 | 5441 | 871 | 1258 |

Significance levels:***p< 0.001,**p < 0.01, *p < 0.05

**Table 3** Authorship patterns and citation impact: negative binomial regression

|  | Model 1 NASU 2020-2021 | Model2 NASU 2022-2023 | Model3 Universities 2020-2021 | Model4 Universities 2022-2023 |
|---|---|---|---|---|
| Intercept | -1.489*** (< 2e-16) | -1.422*** (< 2e-16) | -1.324*** (<2e-16) | -1.226*** (<2e-16) |
| International collaboration | 1.315*** (< 2e-16) | 1.270*** (< 2e-16) | 1.366*** (<2e-16) | 1.313*** (<2e-16) |
| Single author | -0.439*** (3.56e-06) | -0.296** (0.002) | 0.821*** (<2e-16) | -0.508*** (2e-15) |
| Dispersion parameter | 2.8567 | 1.0658 | 1.1206 | 0.6136 |
| Deviance explained | 0.1595 | 0.1319 | 0.1590 | 0.1319 |
| Obs. | 9142 | 8620 | 23099 | 22487 |

Significance levels:***p< 0.001,**p < 0.01, *p < 0.05

Table 1 highlights several key trends in the citation impact of internationally co-authored articles across two periods. Model 1 shows that the citation impact of internationally co-authored articles published by the NASU is unchanged between the two periods. However, Model 2 reveals an increase in the citation impact of internationally co-authored articles published by universities in 2022-2023. Models 3 and 4 demonstrate that internationally co-authored articles published by universities have a citation advantage over those from the NASU, and this advantage has increased in articles published in 2022–2023. Among articles published by the NASU in 2020-2021, the lowest citation impact is observed in mathematics & computer science, while the highest is in social sciences & humanities (Model 5). Among articles published by the NASU in 2022-2023, mathematics & computer science continue to exhibit the lowest impact, followed by physical sciences & engineering (Model 6). Similarly, articles published by universities in 2020–2021 show the lowest citation impact in mathematics & computer science, followed by physical sciences & engineering (Model 7). This trend persists in 2022–2023, with social sciences and humanities achieving the highest citation impact (Model 8).

Table 2 compares the performance of the NASU and universities across disciplines over two periods. In articles published in 2020–2021, the NASU outperforms universities in physical sciences & engineering, as well as, social sciences & humanities. However, in articles published in 2022–2023, the performance gap in physical sciences & engineering has narrowed, and the citation impact in social sciences & humanities becomes statistically equivalent for both groups. Universities consistently outperform the NASU in biomedical & health sciences across both periods, and they exhibit a slight advantage in mathematics & computer science in 2022–2023 (Models 1–5). Although the models explain only a small proportion of variance, this is common in count data, which often display high variability (overdispersion). Such overdispersion can reduce the apparent effect of predictors even when they are significant.

Model 3 specifically examines the relationship between authorship patterns and citation impact. The results show a strong positive effect on FNCI for internationally co-authored articles, while single-authored articles have a negative effect on FNCI compared to domestically co-authored articles.

**Articles in the top 10% most cited globally**

Figure 10 provides insights into the distribution of articles ranked in the top 10% most cited globally classified by authorship pattern: international collaboration, domestic collaboration, and single author. The figure highlights that internationally co-authored articles make up the largest share of articles in the top 10% most cited globally. The share is significantly higher in NASU's research output compared to that of universities. In 2020-2023, the share of internationally co-authored articles in the top 10% most cited globally declined in the research output of both the NASU and universities. The share of domestically co-authored articles published by the NASU increased from 10.8% to 18.6% and the share of single-authored articles published by universities rose from 0.3% to 3.0%.

Figure 11 presents the percentage of articles in the top 10% most cited globally in each category of authorship. In 2020-2023 the share of articles in the top 10% most cited globally fell in the internationally co-authored articles published by the NASU from 7.8% to 6.9%. It increased from 7.0% to 9.6% in the internationally co-authored articles published by universities. The share of the top 10% most cited globally also increased in the domestically co-authored and single-authored articles published by universities. Overall, in 2020-2023, the share of the top 10% most cited globally in the research output of universities increased from 2.6% to 4.6%. In the case of the NASU, it fluctuated around 3.6%, with the percentages of the top 10% most cited globally in domestically co-authored and single-authored articles fluctuating around 0.9% and 0.1% respectively.

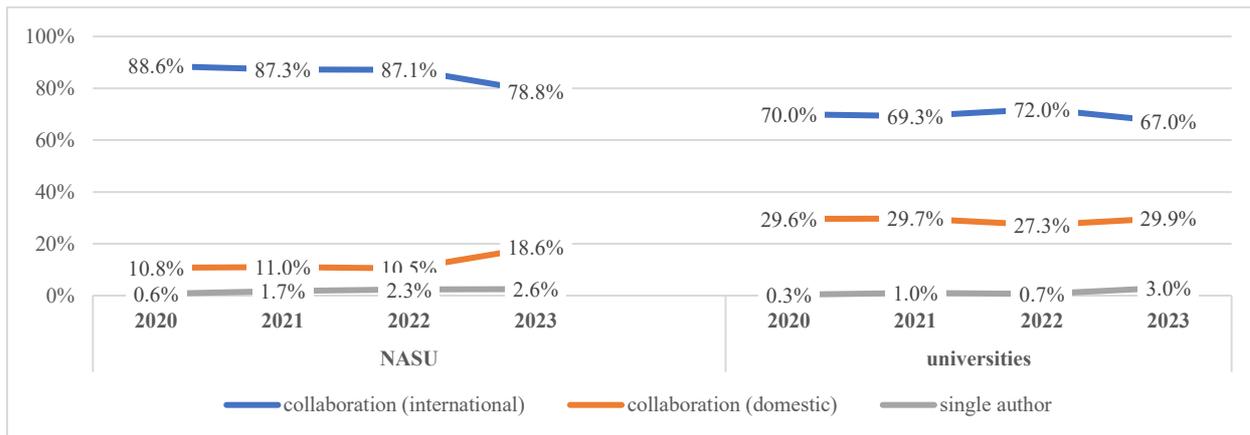

**Fig. 10** Articles in top10% most cited globally classified by authorship pattern

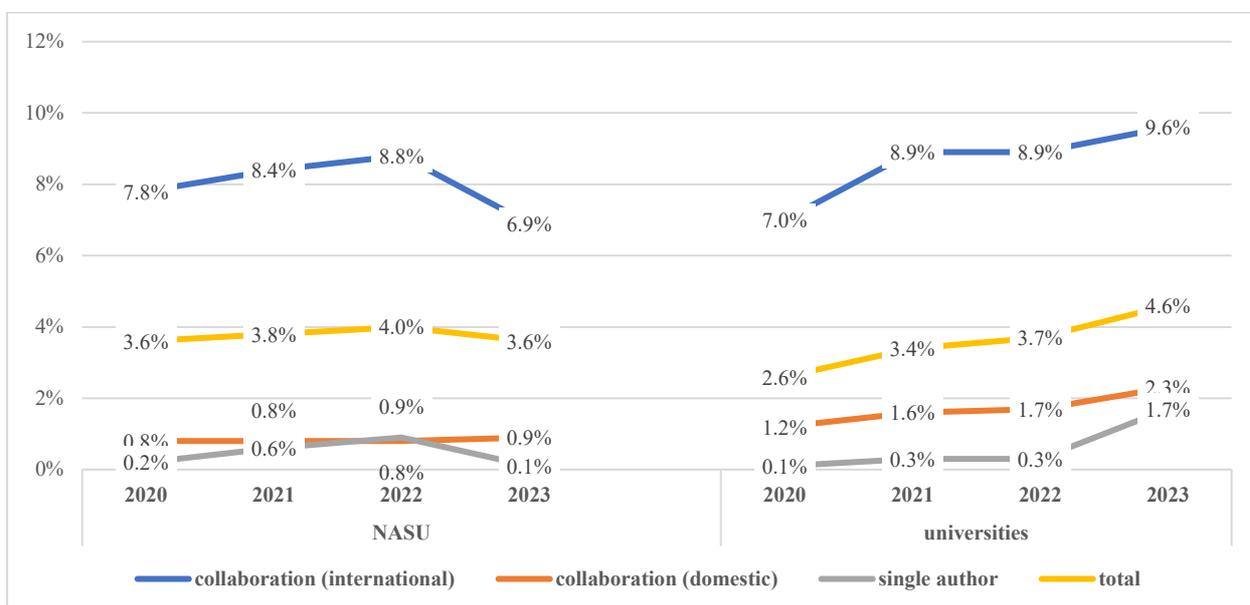

**Fig. 11** The percentage of the top10% most cited globally in each category of authorship

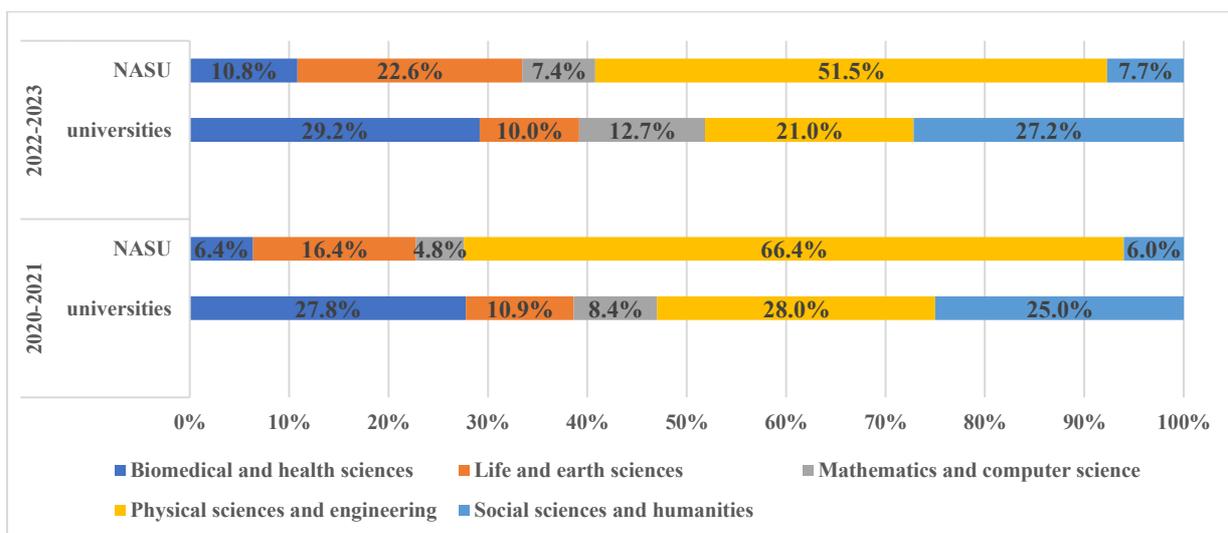

**Fig. 12** Disciplinary prospective on internationally co-authored articles in the top 10% most cited globally

Figure 12 illustrates the distribution of internationally co-authored articles in the top 10% most cited globally across disciplines. It shows that in the case of the NASU, articles in physical sciences & engineering constitute the major share of internationally co-authored articles in the top 10% most cited globally in both periods. They are followed by articles in life & earth sciences. In 2022-2023 the share of articles in physical sciences & engineering increased from 51.5% to 66.4% while the share of articles in life & earth sciences decreased from 22.6% to 16.4%. In the case of universities, the disciplinary distribution of internationally co-authored articles is rather balanced. In 2020-2021 the largest share of internationally co-authored articles published by universities in the top 10% most cited globally constituted articles in biomedical & health sciences. In 2022-2023 articles in physical sciences & engineering took the lead.

**Conclusions**

This study aimed to investigate the implications of Russia's full-scale invasion of Ukraine for the international collaboration of Ukrainian scholars.

Ukrainian scholars deserve respect for continuing to publish despite enduring years of life-threatening conditions, mental strain, shelling, and blackouts.

In terms of countries of international collaboration, it is rather unexpected in 2020-2021, Russia was the top country of international collaboration in articles published by the NASU and the top two in articles published by universities. These articles included both bilateral and multinational collaborations with Russia. Despite the ongoing Russian invasion of Ukraine since 2014 (Oleksiyenko et al., 2021; Oleksiyenko et al., 2023), Ukrainian and foreign scholars did not stop collaborating with Russian scholars. The collaboration of the international academic community with Russian scholars is an indirect support of the Russian invasion of Ukraine.

Poland was the top one country of international collaboration in articles published by universities in both periods and in articles published by the NASU in 2022-2023. These findings resonate with prior studies highlighting a strengthening of collaboration between Ukrainian and Polish scholars in recent years (OECD, 2022). The decisive factor of this collaboration is geographic proximity (Kiselyova & Ivashchenko, 2024). Germany, the USA and China were the other key countries of international collaboration in the articles published by both the NASU and universities. These findings resonate with prior studies highlighting the central role of these countries in global research (Haupt & Lee, 2024; Shih et al., 2024; Zhao, 2024). This prominence is attributed to the financial support mechanisms provided by these countries, which have the highest levels of spending on R&D (Statista, 2024; Sargent, 2022). Ukrainian scholars affiliated with universities intensively collaborated with Slovak scholars and in 2022-2023 their collaboration was strengthened. Results on funding from the collaboration country highlight the significance of funding in promoting international co-authorship (Davies et al., 2022).

In 2022-2023, universities gained more from international collaboration that the NASU. The percentage of internationally co-authored articles remained unchanged for the NASU, while it increased for universities. However, the share of internationally co-authored articles was rather low in the research output of both the NASU and universities. In 2023, it was 40.8 % in articles published by the NASU and 32,2% in articles published by universities. In developed countries, the percentage of internationally co-authored publications is much higher. In 2014 it was 60% in Germany and over 60% in the Netherlands, France and Great Britain (Schmoch et al., 2016).

The citation impact of internationally co-authored articles remained statistically unchanged for the NASU. However, it increased for articles published by universities. The percentage of the articles ranked among the top 10% most cited globally increased in internationally co-authored articles published by universities and fluctuated among those published by the NASU.

The majority of internationally co-authored articles published by the NASU were in physical sciences & engineering in both periods. However, the citation impact of internationally co-authored articles published by the NASU in physical science & engineering in 2022-2023 fell and they almost lost their citation advantage over articles published by universities. Articles published by universities were rather balanced across disciplines, although physical sciences & engineering represented the largest share. In 2022-2023, international collaboration brough gains in biomedical & health sciences and social sciences & humanities for universities. Universities consistently outperformed the NASU in citation impact of internationally co-authored articles in biomedical & health sciences in both periods.

A positive correlation between the citation impact and the number of authors, as well as, foreign affiliation countries was observed in biomedical & health sciences and life & earth sciences. It was more pronounced for universities than the NASU. In social sciences & humanities, the number of authors had minimal effect on citation impact, while in physical sciences & engineering, large collaborations involving the NASU did not correspond to higher FNCI values. Thus the study results only partially support the prior findings that an increase in the number of authors and affiliation countries results in a higher citation impact (Larivière, 2015).

The rather expected finding of the study is that internationally co-authored articles have a higher mean FNCI than domestically co-authored and single-authored articles, supporting the prior studies highlighting that international collaboration results in more impactful publications (Zhou et al., 2020; Persson et al., 2004; Glanzel & Schubert, 2001; Glanzel, 2001; De Moya-Anegon et al., 2018). However, the citation impact differs among disciplines. The smallest advantage in terms **is** citations in internationall co-authored articles is observed in mathematics & computer science for both for NASU and universities. Domestically co-authored and single-authored articles are also among the top 10% most cited globally, though their percentage is much smaller. This indicates that apart from authorship patterns other factors influence the citation impact.

As internationally co-authored articles are more impactful than domestically co-authored and single-authored articles and constitute the major share of the top 10% most cited globally, Ukrainian scholars need to enhance international collaboration. Strengthening international collaboration between Ukrainian scholars and their foreign colleagues requires corresponding financial initiatives and changes in approaches to research funding allocation in Ukraine, as well as, from foreign countries wishing to support Ukrainian scholars in these difficult times. In turn, Ukrainian scholars can contribute to strengthening Europe (Seumenicht, 2024).

**Acknowledgements**
I'm grateful to Rodrigo Costas for insightful and thoughtful comments and suggestions to this study.

**Funding**
This project has received funding through the MSCA4Ukraine project, which is funded by the European Union. Views and opinions expressed are however those of the author only and do not necessarily reflect those of the European Union. Neither the European Union nor the MSCA4Ukraine Consortium as a whole nor any individual member institutions of the MSCA4Ukraine Consortium can be held responsible for them.